# Ultra-High Gradient Channeling Acceleration in Nanostructures: Design/Progress of Proof-of-Concept (POC) Experiments

Y. M. Shin[1,2,a)], A. Green[1], A. H. Lumpkin[2], R. M. Thurman-Keup[2], V. Shiltsev[2]
X. Zhang[3], D. M.-A. Farinella[4], P. Taborek[4], T. Tajima[4]
J. A. Wheeler[5], and G. Mourou[5]

[1]*Northern Illinois Center for Accelerator and Detector Development (NICADD), Department of Physics, Northern Illinois University, Dekalb, IL 60115, USA*
[2]*Accelerator Division, Fermi National Accelerator Laboratory, IL 60510, USA*
[3]*Shanghai Institute of Optics and Fine Mechanics, Shanghai, China*
[4]*University of California - Irvine, Irvine, CA 92697, USA*
[5]*Center for Ultrafast Optical Science and FOCUS Center, University of Michigan, Ann Arbor, Michigan 48109, USA and Laboratoire d' Optique Appliquée, UMR 7639 ENSTA, Ecole Polytechnique, CNRS, Chemin de la Hunière, F-91761 Palaiseau CEDEX, France*

a)Corresponding author: yshin@niu.edu

**Abstract.** A short bunch of relativistic particles or a short-pulse laser perturbs the density state of conduction electrons in a solid crystal and excites wakefields along atomic lattices in a crystal. Under a coupling condition the wakes, if excited, can accelerate channeling particles with TeV/m acceleration gradients [1] in principle since the density of charge carriers (conduction electrons) in solids $n_0 = \sim 10^{20} - 10^{23}$ cm$^{-3}$ is significantly higher than what was considered above in gaseous plasma. Nanostructures have some advantages over crystals for channeling applications of high power beams. The de-channeling rate can be reduced and the beam acceptance increased by the large size of the channels. For beam driven acceleration, a bunch length with a sufficient charge density would need to be in the range of the plasma wavelength to properly excite plasma wakefields, and channeled particle acceleration with the wakefields must occur before the ions in the lattices move beyond the restoring threshold. In the case of the excitation by short laser pulses, the dephasing length is appreciably increased with the larger channel, which enables channeled particles to gain sufficient amounts of energy. This paper describes simulation analyses on beam- and laser (X-ray)-driven accelerations in effective nanotube models obtained from Vsim and EPOCH codes. Experimental setups to detect wakefields are also outlined with accelerator facilities at Fermilab and NIU. In the FAST facility, the electron beamline was successfully commissioned at 50 MeV and it is being upgraded toward higher energies for electron accelerator R&D. The 50 MeV injector beamline of the facility is used for X-ray crystal-channeling radiation with a diamond target. It has been proposed to utilize the same diamond crystal for a channeling acceleration POC test. Another POC experiment is also designed for the NIU accelerator lab with time-resolved electron diffraction. Recently, a stable generation of single-cycle laser pulses with tens of Petawatt power based on thin film compression (TFC) technique has been investigated for target normal sheath acceleration (TNSA) and radiation pressure acceleration (RPA). The experimental plan with a nanometer foil is discussed with an available test facility such as Extreme Light Infrastructure – Nuclear Physics (ELI-NP).

## INTRODUCTION

It has been theoretically postulated that a short bunch of relativistic particles or short pulse lasers perturb the density state of conduction electrons in a solid crystal and excite wakefields along atomic lattices in a crystal. In principle, under a coupling condition the wakes, if excited, can accelerate channeling particles with extremely high acceleration gradients of $G$ (max. gradient) = $m_e c \omega_p / e \approx 96 \times n_0^{1/2}$ [V/m], where $\omega_p = (4\pi n_p e^2 / m_e)^{1/2}$ is the electron plasma frequency and $n_p$ is the ambient plasma density [cm$^{-3}$], $m_e$ and $e$ are the electron mass and charge, respectively, and $c$ is the speed of light in vacuum [1]. A practically obtainable plasma density ($n_p$) in ionized gas is limited to below ~ $10^{18}$ cm$^{-3}$, which in principle corresponds to wakefields up to ~ 100 GV/m. Realistically, it is difficult to create a stable gas plasma with a charge density beyond this limit. Upon high frequency irradiation, metallic crystals can be considered as a naturally existing dense plasma media completely full with a large number of conduction electrons available for the wakefield interactions. The density of charge carriers (conduction electrons) in solids $n_0$ = ~ $10^{20} - 10^{23}$ cm$^{-3}$ is significantly higher than what was considered above in gaseous plasma, and correspondingly the wakefield strength of conduction electrons in solids, if excited, can possibly reach $O$(TV/m) in principle [1 – 7].

Nanostructures have some advantages over the crystals for channeling applications of high power beams. The de-channeling rate can be reduced and the beam acceptance increased by the large size of the channels. For beam driven acceleration, a bunch length with a sufficient charge density would need to be in the range of the plasma wavelength to properly excite plasma wakefields, and channeled particle acceleration with the wakefields must occur before the ions in the lattices move beyond the restoring threshold. In the case of the excitation by short laser pulses of wakefields supported by the electrons in the surrounding nanomaterial wall that is encompassed by the X-ray laser while electrons to be accelerated suffer few collisions from these electrons, while the nanostructure is maintained while laser interact with the material well within the ionization and disintegration times, the dephasing length is appreciably increased with the larger channel, which enables channeled particles to gain a sufficient amount of energy. This paper describes simulation analyses on beam- and laser (X-ray)-driven accelerations in effective CNT models obtained from Vsim [8] and EPOCH [9] codes. Experimental setups to detect wakefields are also outlined with an accelerator facility in Fermilab and NIU.

## PLASMA WAKEFIELD SIMULATIONS OF EFFECTIVE NANOTUBE MODEL

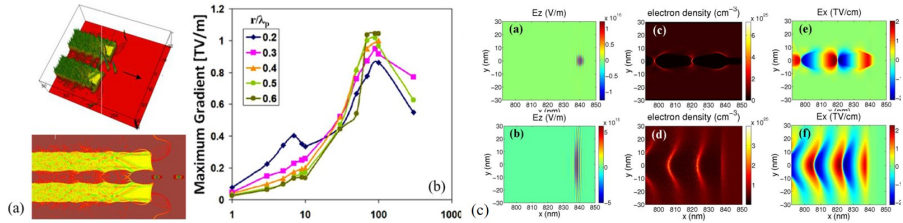

**FIGURE 1.** (a) Charge distribution of a nanotube acceleration ($n_p = n_b$) with a drive and witness beam (b) maximum acceleration gradient versus bunch charge distribution normalized by bunch charge density with various tunnel radii ($r = 0.2 - 0.6\lambda_p$). (c) Wakefield excitation with X-ray laser in a tube (top), in comparison with a wakefield in a uniform system (bottom). (Distributions of (a)(b) the laser field, (c)(d) electron density, and (e)(f) wakefield in terms of a (a, c, e) tube and (b, d, f) uniform density driven by the X-ray pulse).

Figure 1 shows simulation results of crystal wakefield accelerating systems with a hollow plasma channel (effective nanotube) over a uniform plasma that is modeled with $n_p$ ~ $10^{19}$cm$^{-3}$ and ~$10\lambda_p$ length. In the beam-driven acceleration, the simulation condition includes the drive-witness coupling distance of ~$1.6\lambda_p$, $\sigma$ = ~ $0.1\lambda_p$ and a linear regime bunch charge density of $n_b$ ~ $n_p$. For this simulation, the plasma channel is designed with a tunnel of $r = 0.1\lambda_p$. Just like a uniformly filled one, the drive bunch generates tailing wakes in the hollow channel due to the repulsive space charge force. The plasma waves travel along the hollow channel with the density modulation at the same velocity as the witness beam. The traveling wakes around the tunnel continuously transform acceleration energy from the drive beam to the witness one. The energy gain and acceleration gradient are fairly limited by the radius and length of the tunnel with respect to plasma wavelength and bunch charge density. The maximum acceleration gradient is increased from ~ 0.82 TeV/m of $r = 0.2 \lambda_p$ to ~ 1.02 TeV/m of $r = 0.6 \lambda_p$ with $n_b = 100n_p$, corresponding to ~ 20% improvement. In the laser-driven acceleration, the high energy of the photons makes a substantial amount of electrons (either the electrons in the conduction band or in the shallow (< 1 keV) bound electrons) respond to the X-ray fields directly. The high intensity of the X-ray pulse causes the instantaneous ionization of some of the bound electrons per atomic site, thereby contributing to the free electrons. Even some remaining bound electrons may be treated as a solid plasma, where additional optical phonon modes and Buchsbaum resonances are allowed. Two dimensional (2D) particle in cell (PIC) simulations have been performed by using the EPOCH code. For the base case, the laser pulse of wavelength $\lambda_L$ = 1 nm (corresponding to 1 keV X-ray laser) and normalized peak amplitude of $a_0 = 4$ means the peak pulse intensity is $2.2 \times 10^{25}$ W/cm$^2$. The width in the y-direction and length in the x-direction are $\sigma_L = 5\lambda_L$, $\sigma_x = 3\lambda_L$, respectively. The tube wall density is given in terms of the critical density by $n_{tube} = 4.55 \times 10^{-3} n_c$. That is, for modeling the nanotube, a solid tube with wall density of $n_{tube} = 5 \times 10^{24}$/cm$^3$ is used. Figure 1(d) shows the comparison between the nanotube case and uniform density case driven by the X-ray pulse. For the uniform density case (Fig. 1(b, d, f)), the Rayleigh length is short due to the small spot size and so the laser pulse quickly diverges as it propagates. Thus the wakefield becomes weaker and disappears due to the defocusing laser field. In this case, the driving pulse dissipated after propagating a distance of $2000\lambda_L$. However, in the nanotube case (Fig. 2(a, c, e)), the X-ray pulse maintains a small spot size that can be well controlled and guided by the surrounding nanotube walls. The induced wakefield stays stable and the short laser pulse continues propagating even after a distance of $4500\lambda_L$, which is more than twice that of the uniform density case. This stability over a long distance is important for the acceleration to obtain a high energy beam. Considering the real physical parameters, it can be found that the wakefield is higher than 2 TV/cm when driven by the X-ray pulse, which is three orders higher than that of the optical laser case (1 eV laser pulse, $\lambda_L = 1$ μm, the spot size of $\sigma_L = 5$ μm over a length of $\sigma_x = 3$ μm, corresponding to $2.2 \times 10^{19}$ W/cm$^2$, $n_{tube} = 5 \times 10^{18}$/cm$^3$ and tube radius of $\sigma_{tube} = 2.5$ μm).

## OUTLINE OF PLANNED EXPERIMENTS

Proof-of-concept experiments to identify an excitation of solid wakefields are currently being planned and prepared at NIU and Fermilab. CNT targets for the tests will be fabricated by an anodic aluminum oxide (AAO)-CNT template process. The process is an established nanofabrication technique to implant straight, vertical CNTs in a porous aluminum oxide template by the chemical vapor deposition (CVD) growth process. Sub-100 μm long straight multi-wall CNTs grow along the aligned nano-pores in an AAO template, which is followed by thermal cleaning. AAO templates with 100 μm long, 20 – 200 nm wide pores are commercially available. The channeling of the AAO-CNT with $^4$He+ beam was already demonstrated in a low energy regime (2 MeV) by Z. Zhu, et. al. [10].

We first plan a POC experiment with time-resolved electron diffraction analysis at NIU. A sample carbon nanotube diffracts the electrons to an image screen, while being pumped by a short-pulse laser. Relativistic electrons are capable of visualizing temporal dynamics of photo-excited lattices in the area of up to a few micrometer with sub-Angstrom of spatial resolution and sub/pico-second temporal resolution. A fraction of the compressed probe beam is diffracted by atomic scatterers in a thin crystal target or a nanostructured substrate. Diffraction peaks will appear on a screen if the diffraction angle is larger than the divergence of an un-scattered beam and the coherent and elastic scattering intensity is larger than the background noise. If the probe electrons gain energy from an oscillating plasma wave during the scattering process, it will result in a change of diffraction angle deviated from un-pumped one. In principle, an energy gain increases proportional to an interaction length (crystal-target thickness), whereas the thickness is fairly limited by an attenuation of scattered electrons. If the electrons gain a sufficient amount of energy within that distance, the diffraction peak positions will be shifted beyond the screen resolution. Figures 2 (b) and (c) show the pump-probe experimental setup at NIU. A laser pulse is split into two paths, one at IR (800 nm) to pump a sample and another tripled to UV (266 nm) to generate a probe (electron) from a photo-cathode in a RF gun.

Another proposed experiment is planned in the 50 MeV injector beamline of FAST facility [11]. A ~ 3 ps long electron bunch generated from the photo-injector is transported to the magnetic chicane (BC1). The bunch compressed to ~ 1 ps is injected to a CNT target in the goniometer. The channeled electrons are transported to an electron spectrometer and an imaging station (X124). Their energy distribution will be measured by the spectrometer before the beam is dumped to a shielded concrete-enclosure (beam dump). A first beam pre-modulation test was completed with a slit-mask installed in BC1. Preliminary beam-images taken at X124 screen and bunch spectra measured by the MPI indicated a sign of the bunch modulation, but with a low signal-to-noise ratio (Fig. 2(d)). PIC simulations (Vsim) with such density distribution showed that the beam modulation improves a beam-coupling with a crystal target and increased energy gains (Fig. 2(d)). A POC experiment will be set up with scanning beam-injection angles with respect to the target axis and identifying a relative change of projected images from one angle to another, which will be translated into an energy gain/loss of channeled beam.

As an extended effort, an instability-free ion acceleration regime resulting from the interaction of a single cycle pulse and a thin planar foil has been investigated (Fig. 2(e)). This Single-Cycle Laser Acceleration (SCLA) regime permits a thinner optimal target thickness and leads to a more coherent ion layer following the accelerated electron layer. In the present regime, when a single-cycle Gaussian pulse with intensity $10^{23}$W/cm$^2$ is incident on a 50nm planar CH foil, the ponderomotive force of the laser pulse pushes forward an isolated relativistic electron bunch and, in turn, the resultant longitudinal electrostatic field accelerates the protons. With a thin target, our mechanism can coherently and stably accelerate ions over a significant distance without suffering from the typical transverse instabilities that arise under previously considered conditions [12]. This uniquely stable acceleration structure is capable of maintaining a highly monoenergetic ultrashort (~fs) GeV proton bunch. In this way, a compressed ultrashort proton bunch (femtosecond) may be achieved from a standard PW class laser compressed by the Thin Film Compression technique (TFC) [13]. To increase the repetition rate from Hz to kHz, a high-repetition large power laser may be realized by the fiber laser (called CAN laser) [14] in the future. If one combines the CAN laser with the present new technique, we may be also able to access highly repetitive ultrashort proton bunches.

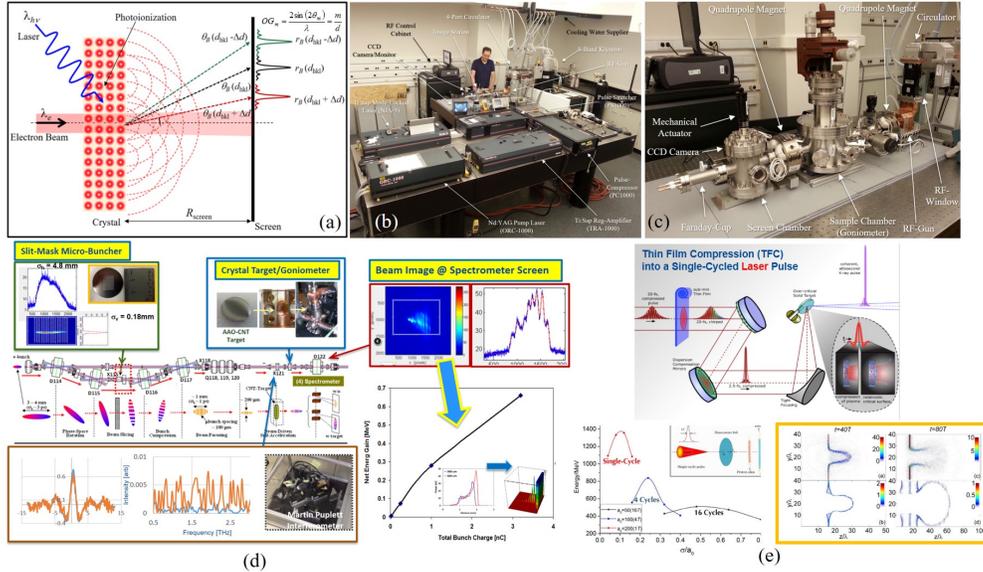

**FIGURE 2.** (a) Conceptual drawing of laser-pumped electron diffraction analysis (b) NIU lab for electron diffraction experiments under construction (c) time-resolved electron diffraction test setup (d) direct energy-shift measurement setup at FAST facility (50 MeV injector beamline), including pre-modulation setup/test results and PIC simulation data (e) thin film compression technique for a single-cycled laser pulse generation (top) and Single-Cycle Laser Acceleration (SCLA, bottom) with PIC simulation data (inset)

## CONCLUSION AND FUTURE WORK

Our PIC simulations on beam- and laser-driven channelling acceleration indicate TeV/m scale accelerating gradients in nanotubes. Two POC experiments are being planned with a test-lab in NIU and in the FAST facility at Fermilab. The SCLA is extensively conceived for high gradient ion accelerations with TFC technique.

## ACKNOWLEDGEMENTS

This work is supported by the DOE contract No. DEAC02-07CH11359 to the Fermi Research Alliance LLC. It is also supported by Extreme Light Infrastructure - Nuclear Physics (ELI-NP), a project co-financed by the Romanian Government and European Union through the European Regional Development Fund.